\documentclass[cits]{PoS}

\graphicspath{{./figs/}}
\usepackage{url}
\usepackage{array}
\usepackage{multirow}
\usepackage{caption}
\usepackage{subcaption}
\usepackage{mdframed}
\usepackage{color}
\usepackage[usenames,dvipsnames]{xcolor}
\usepackage{listings}
\title{Performance of SSE and AVX Instruction Sets }

\ShortTitle{Performance of SSE and AVX Instruction Sets }

\author{\speaker{Hwancheol Jeong}, and Weonjong Lee \\
  Lattice Gauge Theory Research Center, CTP, and FPRD, \\
  Department of Physics and Astronomy, \\
  Seoul National University, Seoul, 151-747, South Korea \\
  E-mail: \email{wlee@snu.ac.kr}}

\author{Sunghoon Kim, and Seok-Ho Myung \\
  Sejong Science High School, Seoul, 152-881, South Korea \\
}

\abstract{ 
SSE (streaming SIMD extensions) and AVX (advanced vector extensions)
are SIMD (single instruction multiple data streams) instruction sets
supported by recent CPUs manufactured in Intel and AMD.
This SIMD programming allows parallel processing by multiple cores in a
single CPU.
Basic arithmetic and data transfer operations such as sum,
multiplication and square root can be processed simultaneously.
Although popular compilers such as GNU compilers and Intel compilers
provide automatic SIMD optimization options, one can obtain better
performance by a manual SIMD programming with proper optimization:
data packing, data reuse and asynchronous data transfer.
In particular, linear algebraic operations of vectors and matrices can
be easily optimized by the SIMD programming.
Typical calculations in lattice gauge theory are composed of linear
algebraic operations of gauge link matrices and fermion vectors, and
so can adopt the manual SIMD programming to improve the performance.
}

\FullConference{The 30th International Symposium on Lattice Field Theory\\
		 June 24 -- 29,  2012\\
		 Cairns, Australia}

\begin{document}

%%%%%%%%%%%%%%%%%%%%%%%%%%%%%%%%%%%%%%%%%%%%%%%%%

\section{Introduction}
SIMD is an abbreviation of the term \emph{Single Instruction, Multiple
  Data streams} \cite{web:SIMD}.
It describes a computer architecture that deals with multiple data
streams simultaneously by a single instruction.
Despite recent CPUs support SIMD instructions, plain C/C++ codes are
composed of SISD (Single Instruction, Single Data streams)
instructions.
However, with SIMD instructions, one can sum multiple numbers
simultaneously, or can calculate a product of vectors with less loops.
In lattice gauge theory, the code are composed of linear algebraic
operations of gauge link matrices and fermion vectors.
Hence, by adopting the SIMD programming such as SSE and AVX, one can
improve the performance of the numerical simulations in lattice gauge
theory \cite{Luscher:2001tx}.
%

%%%%%%%%%%%%%%%%%%%%%%%%%%%%%%%%%%%%%%%%%%%%%%%%%

\section{SIMD Programming\label{sec:simd}}

\begin{figure}[tbhp]
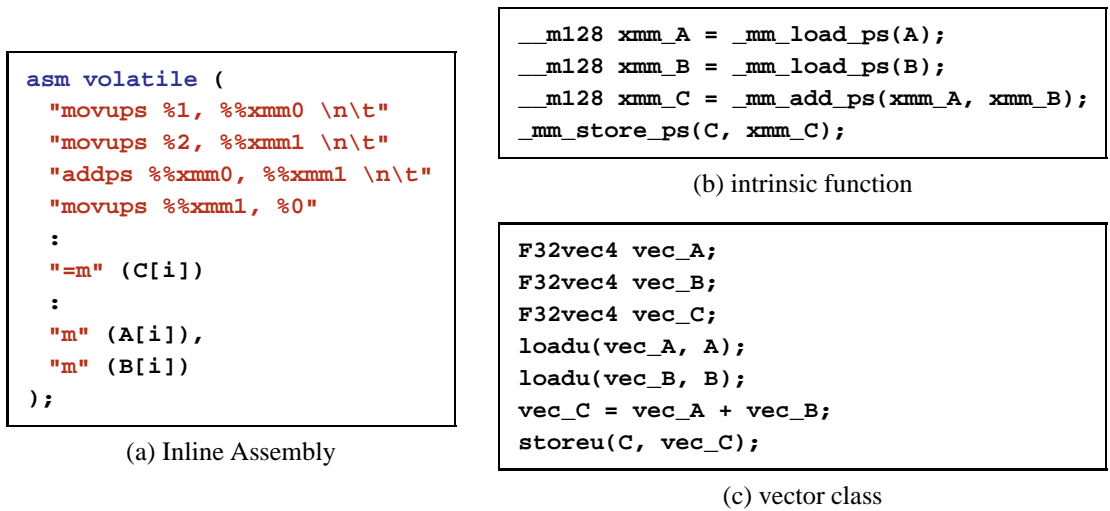

  \centering
  \begin{minipage}{0.4\linewidth}
    \begin{subfigure}[b]{0.9\linewidth}
      \begin{lstlisting}
asm volatile (
  "movups %1, %%xmm0 \n\t"
  "movups %2, %%xmm1 \n\t"
  "addps %%xmm0, %%xmm1 \n\t"
  "movups %%xmm1, %0"
  :
  "=m" (C[i])
  :
  "m" (A[i]),
  "m" (B[i])
);
      \end{lstlisting}
      \vspace{-0.3cm}
      \caption{Inline Assembly}
    \end{subfigure}
  \end{minipage}
  \hspace{0.3cm}
  \begin{minipage}{0.5\linewidth}
    \begin{subfigure}[b]{1\linewidth}
      \begin{lstlisting}
__m128 xmm_A = _mm_load_ps(A);
__m128 xmm_B = _mm_load_ps(B);
__m128 xmm_C = _mm_add_ps(xmm_A, xmm_B);
_mm_store_ps(C, xmm_C);
      \end{lstlisting}
      \vspace{-0.3cm}
      \caption{intrinsic function}
    \end{subfigure}

    \vspace{0.3cm}

    \begin{subfigure}[b]{1\linewidth}
      \begin{lstlisting}
F32vec4 vec_A;
F32vec4 vec_B;
F32vec4 vec_C;
loadu(vec_A, A);
loadu(vec_B, B);
vec_C = vec_A + vec_B;
storeu(C, vec_C);
      \end{lstlisting}
      \vspace{-0.3cm}
      \caption{vector class}
    \end{subfigure}
  \end{minipage}
  \caption{Implementation of SIMD. Codes for the summation of two
    arrays
  }
  \label{fig:simd_method}
\end{figure}

There are three methods to implement the SIMD programming: (1) inline
assembly, (2) intrinsic function, and (3) vector class.
Figure~\ref{fig:simd_method} shows SSE codes that perform the summation of two
arrays using the three methods.
\vspace{-0.7cm}
\begin{itemize}
\item \texttt{Inline assembly}: a merit is that one can handle
  almost every part of a program, so that one can achieve the maximum
  performance of the system.
  One drawback is that writing assembly code is so complicated that it
  requires cautious handling of data transfer between CPUs and
  memories.
\vspace{-0.2cm}
\item \texttt{Intrinsic function}: a merit is that one can program 
it using the standard C/C++ language and so it is easy to program.
A drawback is that there is no guarantee that the code is optimized
to the highest level.
\vspace{-0.2cm}
\item \texttt{Vector class}: a merit is that it is even easier to
  program compared with intrinsic functions.
  A drawback is that the performance is even lower than the intrinsic 
  functions.
\end{itemize}
%
%Figure~\ref{fig:simd_comp} illustrates a comparison between the three
%SIMD implementation methods and C/C++.
%
%\begin{figure}[tbhp]
%  \centering
%  \includegraphics[width=0.5\linewidth]{HowToSIMD}
%  \caption{Performance vs Easiness of Programming}
%  \label{fig:simd_comp}
%\end{figure}
%

%
There have been several SIMD instruction sets for different CPUs.
SSE or AVX are two of them, which are supported by
recent CPUs.
Table~\ref{tbl:proc_dep} shows lists of processors (Intel CPUs) which
support a specific version of SSE and AVX.
The higher version instruction sets include more useful
extensions which are not supported in the lower version.
%
%Show processor dependency
%
\begin{table}[tbhp]
  \centering
  \begin{tabular}{|c|m{13cm}|}
    \hline
    \multirow{4}{*}{\textbf{SSE3}} & Quad-Core Xeon 73xx, 53xx, 32xx,
    \ \ Dual-Core Xeon 72xx, 53xx, 51xx, 30xx, \\
    & Core 2 Extreme 7xxx, 6xxx, \ \ Core 2 Quad 6xxx, \\
    & Core 2 Duo 7xxx, 6xxx, 5xxx, 4xxx, \ \ Core 2 Solo 2xxx, \\
    & Pentium dual-core E2xxx, T23xx \\
    \hline
    \multirow{2}{*}{\textbf{SSE4.1}} & Xeon 74xx, \ \ Quad-Core Xeon
    54xx,
    33xx, \ \ Dual-Core Xeon 52xx, 31xx \\
    & Core 2 Extreme 9xxx, \ \ Core 2 Quad 9xxx, \ \ Core 2 Duo 8xxx,
    Core
    2 Duo E7200 \\
    \hline
    \textbf{SSE4.2} & i7, i5, i3 series, \ \  Xeon 55xx, 56xx, 75xx \\
    \hline
    \textbf{AVX1} & Sandy Bridge, Sandy Bridge-E, Ivy Bridge \\
    \hline
    \textbf{AVX2} & Haswell {\footnotesize(will be released in 2013)}
    \\
    \hline
  \end{tabular}
  \caption{List of SIMD instruction sets for Intel CPUs }
  \label{tbl:proc_dep}
\end{table}

%%%%%%%%%%%%%%%%%%%%%%%%%%%%%%%%%%%%%%%%%%%%%%%%%

%Describe SSE

SSE (Streaming SIMD Extensions) offers SIMD instruction sets for XMM
registers \cite{web:SSE}.
For 64 bit system, there are 16 XMM registers (xmm0 $\sim$ xmm15) in
the CPUs, and an XMM register has 128 bits (16 bytes).
Thus, using XMM registers, one can process four single precision float
point numbers or two double precision floating point numbers
simultaneously.

%Describe AVX

\begin{figure}[tbhp]
  \centering
  \includegraphics[width=0.8\linewidth]{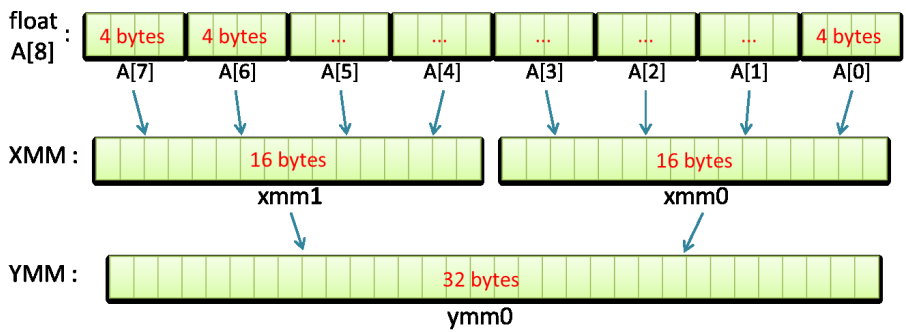}
  \caption{XMM and YMM registers}
  \label{fig:xmm_ymm}
\end{figure}

AVX (Advanced Vector eXtensions) is the next generation of the SIMD
instruction sets supported from Intel Sandy Bridge processors \cite{web:AVX}.
It offers instruction sets for YMM registers.
Similar to the XMM registers, there are 16 YMM registers (ymm0 $\sim$
ymm15) in the CPUS.
The size of ymm is twice bigger than xmm.
Therefore, YMM registers make it possible to process eight single
precision floating point numbers or four double precision floating
point numbers, simultaneously.
Figure~\ref{fig:xmm_ymm} shows a diagram that explains XMM and YMM
registers.
%

%Compare with SSE

Besides the increase in size, AVX also provides an extended
instruction format which allows three input arguments in contrast to
two input arguments allowed for SSE.
Because SSE provides SIMD instructions with only two operand, it is
limited to $a = a + b$ kind of functions.
However, AVX supports three operand SIMD instructions, so that $c = a
+ b$ kind of functions are available using AVX.

%%%%%%%%%%%%%%%%%%%%%%%%%%%%%%%%%%%%%%%%%%%%%%%%%

%%%%%%%%%%%%%%%%%%%%%%%%%%%%%%%%%%%%%%%%%%%%%%%%%

\section{Optimization Scheme}

%%%%%%%%%%%%%%%%%%%%%%%%%%%%%%%%%%%%%%%%%%%%%%%%%

\subsection{Data Packing\label{subsec:packing}}

%Describe data packing algorithm

As described in Section~\ref{sec:simd}, an advantage of SIMD
programming is the data packing.
One can pack multiple data to a single XMM or YMM register and those
data can be processed or calculated simultaneously using SSE or AVX
instructions.
Recent SSE and AVX instructions provide many useful SIMD functions --
sum, multiply, square root, shift, \textit{etc.}

%Show examples (sum, data copy)
%Show its limit (Compiler's automatic optimization)
%
\begin{figure}[tbhp]
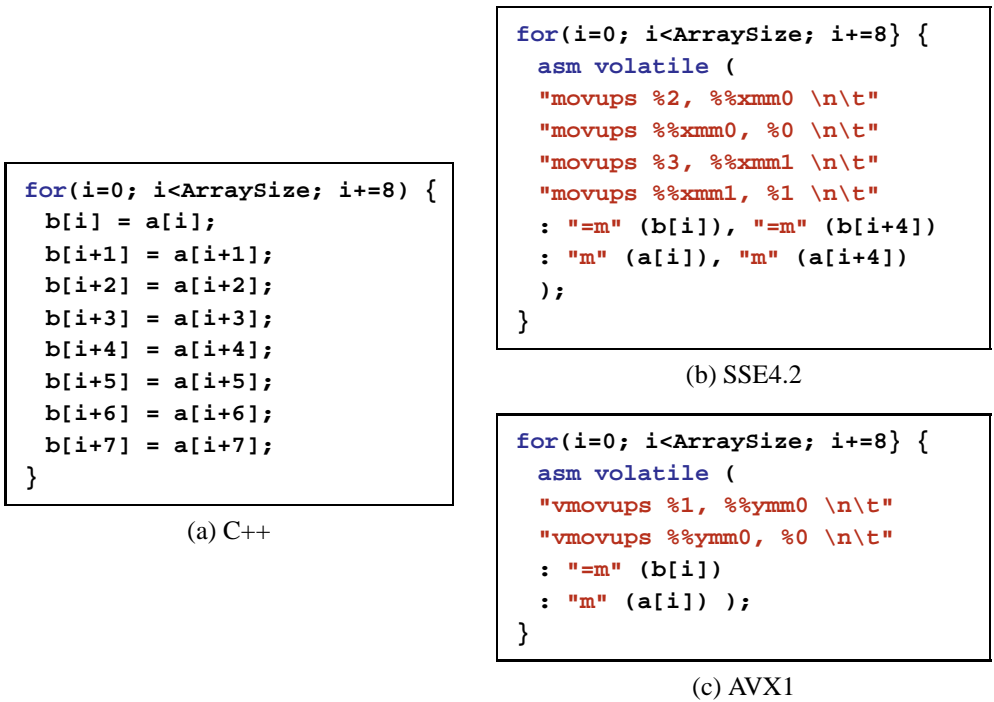

  \centering
  \begin{minipage}{0.4\linewidth}
    \begin{subfigure}[b]{0.9\linewidth}
      \begin{lstlisting}
for(i=0; i<ArraySize; i+=8) {
  b[i] = a[i];
  b[i+1] = a[i+1];
  b[i+2] = a[i+2];
  b[i+3] = a[i+3];
  b[i+4] = a[i+4];
  b[i+5] = a[i+5];
  b[i+6] = a[i+6];
  b[i+7] = a[i+7];
}
      \end{lstlisting}
      \vspace{-0.3cm}
      \caption{C++}
    \end{subfigure}
  \end{minipage}
  \hspace{0.3cm}
  \begin{minipage}{0.4\linewidth}
    \begin{subfigure}[b]{1\linewidth}
      \begin{lstlisting}
for(i=0; i<ArraySize; i+=8} {
  asm volatile (
  "movups %2, %%xmm0 \n\t"
  "movups %%xmm0, %0 \n\t"
  "movups %3, %%xmm1 \n\t"
  "movups %%xmm1, %1 \n\t"
  : "=m" (b[i]), "=m" (b[i+4])
  : "m" (a[i]), "m" (a[i+4])
  );
}
      \end{lstlisting}
      \vspace{-0.3cm}
      \caption{SSE4.2}
    \end{subfigure}

    \vspace{0.3cm}

    \begin{subfigure}[b]{1\linewidth}
      \begin{lstlisting}
for(i=0; i<ArraySize; i+=8} {
  asm volatile (
  "vmovups %1, %%ymm0 \n\t"
  "vmovups %%ymm0, %0 \n\t"
  : "=m" (b[i])
  : "m" (a[i]) );
}
      \end{lstlisting}
      \vspace{-0.3cm}
      \caption{AVX1}
    \end{subfigure}
  \end{minipage}
  \caption{\label{fig:data_copy} Codes for the simple data copy made
    in (a) C++, (b) SSE4.2, and (c) AVX1.}
\end{figure}
Figure~\ref{fig:data_copy} shows partial codes of a program which
performs a simple data copy between two arrays, written in
plain\footnote{Here \emph{plain} C/C++ denotes a normal C/C++ language
  without SSE or AVX. Since SSE or AVX are also implemented in C/C++ 
  code, we use \emph{plain} to distinguish those programming methods
  from the normal C/C++ language.}  C++, SSE4.2 and AVX1 language
respectively.
For SIMD code in the figure (SSE and AVX), the data copy code is
implemented using inline assembly method in order to obtain maximum
performance.

Table~\ref{tbl:data_copy} shows the performances of the three
different methods when the size of the array is $10^9$.
Without optimization option for compiler, SIMD methods (SSE4.2 and
AVX1) are much faster than plain C++ code.
When the maximum optimization option is applied to the C++ compiler,
the SIMD method is as fast as C++.
This result indicates that the compiler optimizes the given C++ code
by converting to SIMD code, automatically.
Indeed, we can convert the C++ object code into an assembler code
using the \texttt{objdump} command in LINUX, and then we can have a
look at the assembler source code.
In this way, we find out that the optimization option convert the
C++ code into an assembler code using the SIMD instruction sets
automatically.

We also find out that, regardless of the optimization, performances of
SSE4.2 and AVX1 are almost the same.
This indicates that use of AVX1, i.e., YMM registers, does not improve
the speed of data transfer.
\begin{table}[tbhp]
  \centering
  \begin{tabular}{| >{\centering}m{2cm} | >{\centering}m{1.5cm} |
      >{\centering}m{1.5cm} | >{\centering}m{1.5cm}|}
    \hline
    & C++ & SSE4.2 & AVX1 \tabularnewline
    \hline
    no opt. & 163 & 94.5 & 97.7 \tabularnewline
    max. opt. & 75 & 71 & 75 \tabularnewline
    \hline
  \end{tabular}
  \caption{\label{tbl:data_copy}
    CPU clocks required for simple data transfer of array of $10^9$ size, 
    using
    the codes given in Figure~\protect\ref{fig:data_copy} in units of
    $10^4$ clocks. Here `no opt.' corresponds to the results that the
    optimization option is turned off with the C++ compiler. The `max. opt.'
    corresponds to the results with the maximum 
    optimization option turned on. 
    We use Intel Core i7-3820 Sandy
    Bridge-E with Fedora 17 of kernel 3.5.2-3.fc17. The compiler is GCC
    4.7.0.
  }
\end{table}
%

%%%%%%%%%%%%%%%%%%%%%%%%%%%%%%%%%%%%%%%%%%%%%%%%%

\subsection{Data Reuse}

%Describe data reuse algorithm

In the previous example of data transfer, we find that the C++
optimization option makes the program to use the SIMD instructions.
However, one can increase the performance of the SIMD code using data
reuse.
In the inline assembly, a programmer has a full control over
registers.
Thus, if there are some data which are repeatedly used, one can reuse
existing register, so that one may remove unwanted data transfer.

%Show example (repeated sum)
%
\begin{figure}[tbhp]
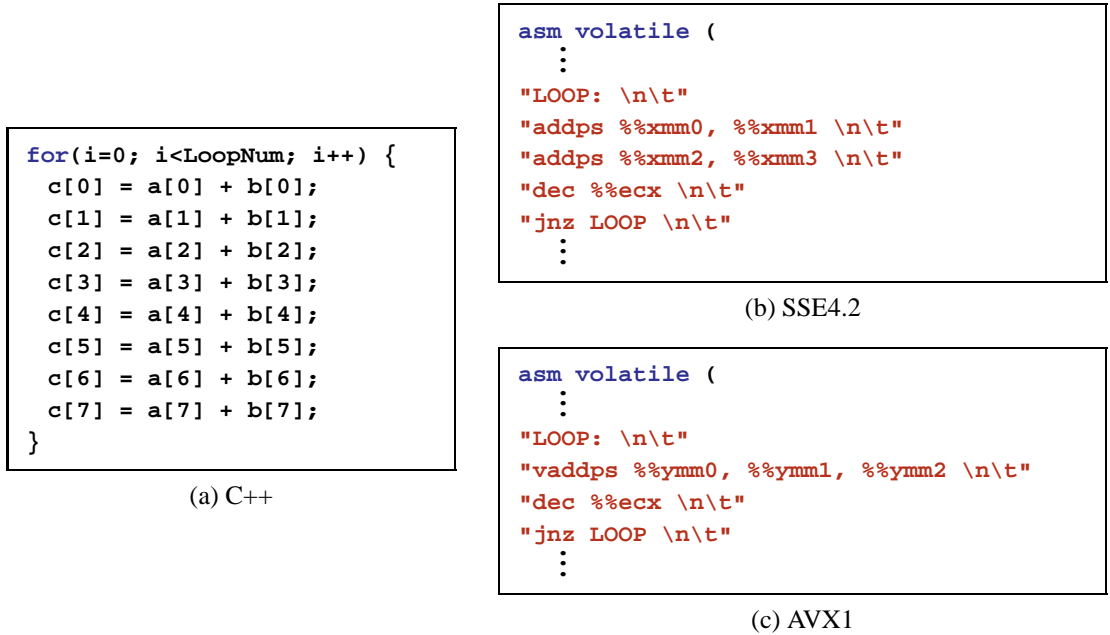

  \centering
  \begin{minipage}{0.4\linewidth}
    \begin{subfigure}[b]{0.9\linewidth}
      \begin{lstlisting}
for(i=0; i<LoopNum; i++) {
  c[0] = a[0] + b[0];
  c[1] = a[1] + b[1];
  c[2] = a[2] + b[2];
  c[3] = a[3] + b[3];
  c[4] = a[4] + b[4];
  c[5] = a[5] + b[5];
  c[6] = a[6] + b[6];
  c[7] = a[7] + b[7];
}
      \end{lstlisting}
      \vspace{-0.3cm}
      \caption{C++}
    \end{subfigure}
  \end{minipage}
  \hspace{0.3cm}
  \begin{minipage}{0.5\linewidth}
    \begin{subfigure}[b]{1\linewidth}
      \begin{lstlisting}
asm volatile (
|\hspace{0.5cm}\raisebox{-1pt}[0pt][0pt]{$\vdots$}|
"LOOP: \n\t"
"addps %%xmm0, %%xmm1 \n\t"
"addps %%xmm2, %%xmm3 \n\t"
"dec %%ecx \n\t"
"jnz LOOP \n\t"
|\hspace{0.5cm}\raisebox{-1pt}[0pt][0pt]{$\vdots$}|
      \end{lstlisting}
      \vspace{-0.3cm}
      \caption{SSE4.2}
    \end{subfigure}

    \vspace{0.3cm}
    
    \begin{subfigure}[b]{1\linewidth}
      \begin{lstlisting}
asm volatile (
|\hspace{0.5cm}\raisebox{-1pt}[0pt][0pt]{$\vdots$}|
"LOOP: \n\t"
"vaddps %%ymm0, %%ymm1, %%ymm2 \n\t"
"dec %%ecx \n\t"
"jnz LOOP \n\t"
|\hspace{0.5cm}\raisebox{-1pt}[0pt][0pt]{$\vdots$}|
      \end{lstlisting}
      \vspace{-0.3cm}
      \caption{AVX1}
    \end{subfigure}
  \end{minipage}
  \caption{\label{fig:simple_sum} Codes for simple summation made in
    (a) C++, (b) SSE4.2, and (c) AVX1.}
\end{figure}
Figure~\ref{fig:simple_sum} shows part of the codes that does
a simple summation, written in C++, SSE4.2, and AVX1.
To illustrate the advantage of data reuse, we repeat summations many
times.
Since the same data are used repeatedly, the SIMD codes (SSE4.2 and
AVX1) remove unwanted data transfer of reloading the same variables,
and reuse the existing data in registers.

Table~\ref{tbl:simple_sum} compares the performances of the three
methods by performing a simple summation $10^9$ times over the same
data.
We find out that regardless of the compiler optimization, the SIMD
methods (SSE4.2 and AVX1) are significantly faster by an order of
magnitude than C++.
Furthermore, AVX1 is much faster (almost 3 times) than SSE4.2.
This result indicates that by adjusting the AI (arithmetic intensity)
ratio\footnote{Here, we use the standard definition of AI, which is
  the ratio of the amount of floating point calculation and the amount
  of data transfer.} using such an optimization method as data reuse
and trading the data transfer with the floating point calculation by
SU(3) reconstruction, we can increase the performance of the SIMD
programming dramatically.
\begin{table}[tbhp]
  \centering
  \begin{tabular}{| >{\centering}m{2cm} | >{\centering}m{1.5cm} |
      >{\centering}m{1.5cm} | >{\centering}m{1.5cm}|}
    \hline
    & C++ & SSE4.2 & AVX1 \tabularnewline
    \hline
    no opt. & 732 & 83.7 & 26.9 \tabularnewline
    max. opt. & 317 & 78.8 & 26.9 \tabularnewline
    \hline
  \end{tabular}
  \caption{\label{tbl:simple_sum} 
    CPU clocks required for a simple summation using the
    codes given in Figure~\protect\ref{fig:simple_sum} in units of $10^4$
    clocks. The index convention is the same as in Table 
    \protect\ref{tbl:data_copy}. 
  }
\end{table}
%

%%%%%%%%%%%%%%%%%%%%%%%%%%%%%%%%%%%%%%%%%%%%%%%%%

\subsection{Asynchronous Data Transfer}

%Describe asynchronous data transfer

Data transfer from the data memory to the registers is a slow process.
As discussed in Subsection~\ref{subsec:packing}, the gain of using XMM
and YMM registers for the data transfer is only a factor of 1.2.
Hence, in a real code, it is hard to obtain the full advantage of
the data packing.
Fortunately, the overload of the data transfer can be minimized by
using the asynchronous data transfer method.
Asynchronous data transfer is a technique that performs calculation
and data transfer, simultaneously.

%Explain prefetching method
%
SSE and AVX instructions provide some basic prefetching methods.
Prefetching is a technique which pre-loads data to the cache memory,
before CPU initiates the calculation.
However, the prefetching in SSE and AVX does not support a full
control over the memory caching, but only gives hints to CPU about
the memory caching.
In other words, it does not force data to be preloaded to the cache
memory, but just give information on which data hope to be pre-loaded.
This prefetching method does not work successfully because there are
many background processes from OS or other applications to be handled
with higher priority from the standpoint of CPU.

%
%Show prefetching example
%(Further improvement is under investigation)
%
\begin{figure}[tbhp]
  \centering
  \begin{subfigure}[b]{0.45\linewidth}
    \begin{lstlisting}
|\hspace{0.5cm}\raisebox{-1pt}[0pt][0pt]{$\vdots$}|
"LOOP: \n\t"
|{\color{ForestGreen}"prefetch1 0x200(\%\%rax)\ 
  {\@backslashchar}n{\@backslashchar}t"}|
"movups (%%rax), %%xmm0 \n\t"
"movups %%xmm0, (%%rdx) \n\t"
"movups 0x10(%%rax), %%xmm1 \n\t"
"movups %%xmm1, 0x10(%%rdx) \n\t"
|\hspace{0.5cm}\raisebox{-1pt}[0pt][0pt]{$\vdots$}|
    \end{lstlisting}
    \vspace{-0.3cm}
    \caption{SSE4.2}
  \end{subfigure}
  \hspace{0.5cm}
  \begin{subfigure}[b]{0.45\linewidth}
    \begin{lstlisting}
|\hspace{0.5cm}\raisebox{-1pt}[0pt][0pt]{$\vdots$}|
"LOOP: \n\t"
|{\color{ForestGreen}"prefetch1 0x200(\%\%rax)\ 
  {\@backslashchar}n{\@backslashchar}t"}|
"movups (%%rax), %%ymm0 \n\t"
"movups %%ymm0, (%%rdx) \n\t"
"movups 0x20(%%rax), %%ymm1 \n\t"
"movups %%ymm1, 0x20(%%rdx) \n\t"
|\hspace{0.5cm}\raisebox{-1pt}[0pt][0pt]{$\vdots$}|
    \end{lstlisting}
    \vspace{-0.3cm}
    \caption{AVX1}
  \end{subfigure}
  \caption{\label{fig:prefetch} Codes for data copy using prefetching
    techniques.}
\end{figure}
Figure~\ref{fig:prefetch} shows SIMD codes of data transfer with
prefetching method.
The results presented in Table~\ref{tbl:prefetch} shows that the
prefetching method improves about 1.5\% for SSE and 5\% for AVX method.
Hence, we need a more powerful asynchronous data transfer method.
\begin{table}[tbhp]
  \centering
  \begin{tabular}{| >{\centering}m{4cm} | >{\centering}m{1.5cm} |
      >{\centering}m{1.5cm} | >{\centering}m{1.5cm}|}
    \hline
    & C++ & SSE4.2 & AVX1 \tabularnewline
    \hline
    no prefetching & 75 & 71 & 75 \tabularnewline
    512 bytes prefetching & & 69.9 & 70.9 \tabularnewline
    \hline
  \end{tabular}
  \caption{\label{tbl:prefetch}
    CPU clocks required for data transfer of array of $10^9$ single
    precision floating
    point numbers with
    prefetching method. We use the codes in
    Figure~\protect\ref{fig:prefetch}. Units are $10^4$ clocks.
  }
\end{table}

%%%%%%%%%%%%%%%%%%%%%%%%%%%%%%%%%%%%%%%%%%%%%%%%%

%%%%%%%%%%%%%%%%%%%%%%%%%%%%%%%%%%%%%%%%%%%%%%%%%

%%%\input{application.tex}

%%%%%%%%%%%%%%%%%%%%%%%%%%%%%%%%%%%%%%%%%%%%%%%%%

\section{Conclusion}

%Conclusion

Recent CPUs support the SIMD instructions such as SSE and AVX.
The SIMD instruction sets provide methods of parallel processing in a
single CPU level.
The standard C/C++ compilers support those SIMD instructions with
optimization options.
It turns out that one can achieve significantly higher performance by
programming SIMD codes using the inline assembly as demonstrated in
this paper.
The optimization techniques of SIMD programming such as data packing,
data reuse, and asynchronous data transfer can be easily applied to
the physics code in lattice gauge theory.
%

%%%%%%%%%%%%%%%%%%%%%%%%%%%%%%%%%%%%%%%%%%%%%%%%%

\section{Acknowledgement}

%C.~Jung is supported by the US DOE under contract DE-AC02-98CH10886.
%
The research of W.~Lee is supported by the Creative Research
Initiatives Program (2012-0000241) of the NRF grant funded by the
Korean government (MEST).
W.~Lee would like to acknowledge the support from the KISTI
supercomputing center through the strategic support program for the
supercomputing application research [No. KSC-2011-G2-06].
%
% The work of S.~Sharpe is supported in part by the US DOE grant
% no.~DE-FG02-96ER40956.
%
Computations were carried out in part on QCDOC computing facilities of
the USQCD Collaboration at Brookhaven National Lab, on GPU computing
facilities at Jefferson Lab, on the DAVID GPU clusters at Seoul
National University, and on the KISTI supercomputers. The USQCD
Collaboration are funded by the Office of Science of the
U.S. Department of Energy.

%%%%%%%%%%%%%%%%%%%%%%%%%%%%%%%%%%%%%%%%%%%%%%%%%

%%%%%%%%%%%%%%%%%%%%%%%%%%%%%%%%%%%%%%%%%%%%%%%%%

\end{document}